\documentclass[
reprint,
 amsmath,amssymb,
 aps,
 prd,
 floatfix
]{revtex4-2}

\usepackage{graphicx}
\usepackage{subcaption}
\usepackage{dcolumn}
\usepackage{bm}

\usepackage{slashed}
\usepackage{esvect,esint}
\usepackage{braket}
\usepackage{xcolor}

\usepackage{hyperref}

\newcommand{\dd}[1]{\mathrm{d}#1}

\begin{document}

\preprint{APS/123-QED}

\title{
	\texorpdfstring{Extracting Nucleon Resonance Transition GPDs from \(e^- N\to e^-\gamma N\pi\) Deeply Virtual Compton Scattering}
	{Extracting Nucleon Resonance Transition GPDs from e N -> e gamma N pi Deeply Virtual Compton Scattering}
}

\author{Matthew Rumley}
\email{matthew.rumley@adelaide.edu.au}

\author{Anthony W. Thomas}
\affiliation{Adelaide University}

\collaboration{Centre for the Subatomic Structure of Matter}

\date{\today}

\begin{abstract}
We investigate the process in which Deeply Virtual Compton Scattering (DVCS) excites a baryon resonance. In particular, we assess, in DVCS leading to the Roper resonance, the relative importance of a ``background'' process in which a pion is first emitted by the nucleon, which then undergoes a DVCS event. Our numerical results, using realistic DVCS kinematics, indicate that there can be measurable interference effects. They suggest that this process could substantially modify the experimentally observed cross sections at CLAS12-like kinematics, motivating their inclusion in precision analyses of DVCS experiments. We further find that in spite of this background, the transition to a Roper-like state through DVCS does contribute significantly to the \(e^- N\to e^-\gamma N\pi\) cross section in some kinematic regions. This suggests that the creation of nucleon resonances via DVCS is a useful method for extracting information about the nucleon transition GPDs and the internal structure of the excited states.
\end{abstract}

\maketitle

\section{Introduction}
Generalised Parton Distributions (GPDs) provide a useful framework for studying the dynamical structure of hadrons through hard, exclusive processes such as Deeply Virtual Compton Scattering (DVCS) \cite{Ji:1996ek,Diehl,BELITSKY20051}. Through such studies of internal hadron structure, we expect to gain new insight into the fundamentals of Quantum Chromodynamics (QCD) and thus strong interaction phenomena.

GPDs encode a transverse femtographic image of the internal structure of the hadron which, along with the usual longitudinal momentum information found in a parton distribution function (PDF), form a complete picture of the parton structure of hadrons~\cite{Burkardt:2000za,Thomas:2025rht,Boer:2025ixc}. Here, we consider the prospect of extracting \(N\rightarrow N^*\) transition GPDs through non-diagonal DVCS processes such as \(N \rightarrow N\pi\) where the final \(N\pi\) state resonates at 1440 MeV, the invariant mass of the \(P_{11}\) (Roper) resonance~\cite{Roper:1964zza}. Through this we can further our understanding of how QCD dynamics is involved in the formation of resonances. This is especially important in the case of the Roper resonance, where there is considerable disagreement on whether it should be understood as a genuine three-quark excited state~\cite{Burkert:2017djo}, or whether it is dynamically generated by strong meson-baryon scattering~\cite{Wu:2017qve,Owa:2025mep}. 

In this paper, we closely follow the formalism found in Semenov {\em et al.}~\cite{Semenov-Tian-Shansky:2023bsy} and calculate the cross section for a \(e^- N \rightarrow e^- \gamma P_{11}(1440) \rightarrow e^- \gamma N \pi\) DVCS process. However, in addition, we consider the effect of adding the process, which we label \textbf{``diagonal''}, in which the nucleon first emits a pion and then undergoes DVCS.
We detail a phenomenological model for both the transition GPDs and a compatible set of diagonal GPDs. Then we calculate the scattering amplitudes before and after the addition of the background process, and discuss the impact of the background process on our ability to extract transition GPDs from this resonance-producing process. 
These pion emission channels have typically been neglected in previous DVCS analyses \cite{HERMES:2001bob,CLAS:2001wjj,H1Collaboration,JeffersonLabHallA:2022pnx}. However, their interference with resonance-mediated DVCS could introduce significant corrections to the cross sections, potentially affecting GPD extractions from experimental data.

The full electroproduction cross section receives contributions from the Bethe--Heitler mechanism, the DVCS mechanism, and their mutual interference \cite{Diehl,BELITSKY20051,Muller}. Although the Bethe--Heitler contribution is of course crucial, our aim is methodological. We present a sensitivity study of transition-GPD-driven structure in the hadronic DVCS amplitude, concentrating on the structure of the modified DVCS amplitude and the associated final–state interactions.  A quantitative phenomenological analysis including the Bethe--Heitler channel and interference terms is postponed to future studies. 

In Section~\ref{sec:kinematics} we introduce the kinematics of the reaction. Section~\ref{sec:scatteringAmplitude} details the scattering amplitude for both the transition and diagonal processes, including the parameterization of a set of transition and nucleon GPDs. We then create a model for these GPDs in Section~\ref{sec:GPDModels}. In Section~\ref{sec:analysis} we present the results for the cross sections and discuss the effect of including the background process on our ability to obtain transition GPDs. Finally, our conclusions and outlook for future work are presented in Section \ref{sec:conclusion}.

\section{Kinematics} \label{sec:kinematics}
We label the momenta of the hard, exclusive DVCS process as follows:
\begin{gather}
    e^-(k_e) + N(p) \rightarrow e^-(k_e') + \gamma(q') + N(p_f) + \pi(k_\pi).
\end{gather}

For the non-diagonal DVCS process the intermediate step takes the form
\begin{align}
    \nonumber e^-(k_e) + N(p) & \rightarrow e^-(k_e') + \gamma(q') + P_{11}(p_R) \\
    & \rightarrow e^-(k_e') + \gamma(q') + N(p_R - k_\pi) + \pi(k_\pi).
\end{align}

For the diagonal background process this is
\begin{align}
    \nonumber e^-(k_e) + N(p) & \rightarrow e^-(k_e') + I(p-k_\pi) + \pi(k_\pi) \\
    & \rightarrow e^-(k_e') + \gamma(q') + N(p_R-k_\pi) + \pi(k_\pi) \, ,
\end{align}
where \(I\) represents the intermediate off-shell state after the pion is emitted.

We then create a series of kinematic variables in a manner similar to most DVCS setups \cite{Semenov-Tian-Shansky:2023bsy,BELITSKY20051}
\begin{gather}
    \nonumber q = k_e - k_e',\quad \Delta=p_R-p \\
    \nonumber Q^2=-q^2,\quad x_B=\frac{Q^2}{2p\cdot q},\quad y=\frac{p\cdot q}{p\cdot k_e} \\
    \nonumber t = \Delta^2,\quad M_{\pi N}^2=M_R^2=p_R^2,\quad \Phi,\quad \theta_\gamma,\quad  \theta_\pi^*,\quad \phi_\pi^* \, ,
\end{gather}
where \(\Phi\) is the angle between the leptonic and hadronic planes (defined by \(\vv{k_e}\times \vv{k'_e}\) and \(\vv{q}\times \vv{q'}\) respectively) as defined in the lab frame (initial nucleon three-momentum \(\vv{p}=0\)). The angle \(\theta_\gamma\) is the angle the real photon is emitted at relative to the incoming beam in this frame. The angles \(\theta_\pi^*\) and \(\phi_\pi^*\) indicate the polar and azimuthal angles of the final-state pion in the rest frame of the \(N\pi\) system (also the rest frame of the intermediate resonance) taken relative to \(\vv{p_R}\). 

\begin{figure}
    \centering
    \includegraphics[width=\linewidth]{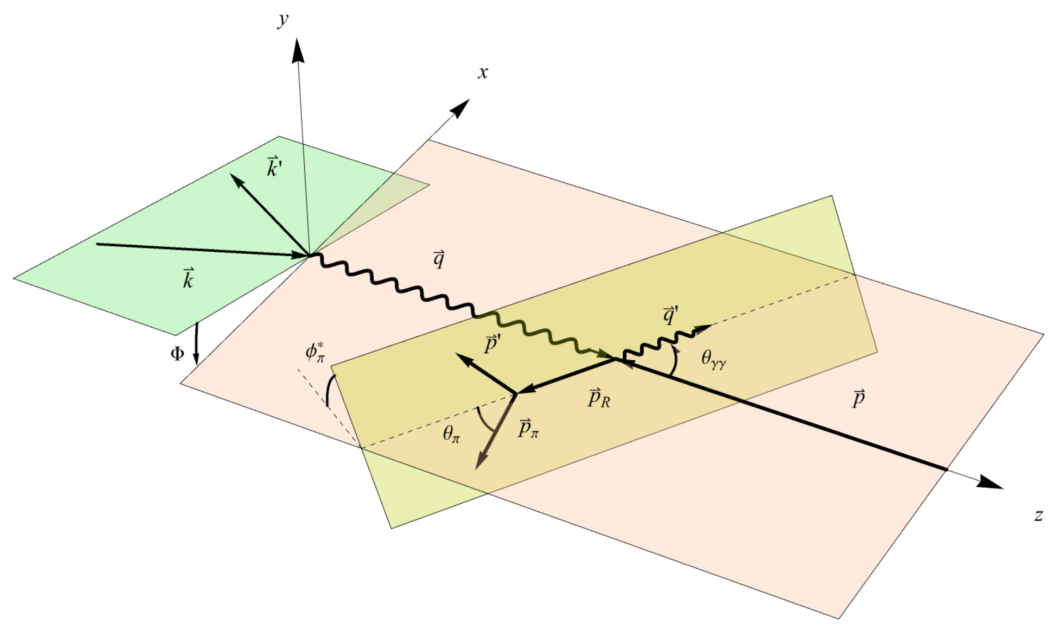}
    \caption{A schematic for the kinematics of the DVCS to excited state process \cite{Semenov-Tian-Shansky:2023bsy}.}
    \label{fig:kinematics}
\end{figure}

We will use the convention that all variables are taken in the lab frame except where indicated by a superscript, * , which corresponds to the \(N\pi\) rest frame.

The 7-fold differential cross section for this process is
\begin{align}
    \nonumber \frac{\dd\sigma}{\dd Q^2 \dd x_B \dd t \dd \Phi \dd M_{\pi N}^2 \dd\Omega_\pi^*} &=  \frac{1}{(2\pi)^7}  \frac{x_B\,y^2}{32Q^4\sqrt{1+\dfrac{4M_N^2 x_B^2}{Q^2}}} \\
    \times \frac{|\vec k^*_\pi|}{4 M_{\pi N}} & \sum_{\{\text{spins}\}} \left| \mathcal{M}(eN\!\to\!e\gamma\pi N)\right|^2 ,
\end{align}
where the sum represents the average (sum) over initial (final) state spins and helicities and
\begin{align}
    \nonumber |\vv{k^*_\pi}|=\frac{1}{2M_{\pi N}}\lambda^{1/2}\!\left(M_{\pi N}^2, \, M_N^2, \, m_\pi^2\right) \, ,
\end{align}
with \(\lambda(x,y,z)\) the K{\"a}ll{\'e}n triangle function~\cite{Semenov-Tian-Shansky:2023bsy}.

We move to plotting a 5-fold differential cross section against the invariant mass of the \(N\pi\) system by applying the transformation
\begin{align}
    \dd M_{\pi N}^2 = 2M_{\pi N}\,\, \dd M_{\pi N} \, ,
\end{align}
integrating over the pion decay solid angle \(\dd\Omega^*_\pi\), and evaluating the cross section at fixed values of \(Q^2,\,x_B,\,t,\,\Phi\). Of course, as mentioned earlier, this cross section is purely for illustration, with inclusion of the interference with the Bethe-Heitler process essential for comparison with experiment.

\section{Scattering Amplitudes} \label{sec:scatteringAmplitude}
\subsection{\texorpdfstring{Transition to a $1/2^+$ Resonance}{Transition to a 1/2+ Resonance}} \label{sec:transitionAmplitude}
The transition amplitude for the case where a target nucleon is struck by a virtual photon, emits a real photon, and transitions to the Roper resonance (\(J^P = \frac{1}{2}^+\)), which then decays into \(N\pi\), is given by
\begin{widetext}
\begin{equation} 
    i\mathcal{M} = \frac{e\,g_f\,g_A\,\tau^a}{2f_\pi} \, \bar{u}_e(k_e',s') \, \gamma_\mu \, u_e(k_e,s) \, \varepsilon^*_\nu(q',\lambda') \, \bar{u}_N(p_R-k_\pi,s_N') \, \slashed{k}_\pi\gamma^5 \, S_f(p_R,M_R) \, H^{\mu\nu}_{NP_{11}} \, u_N(p,s_N) \, ,
\end{equation}
\end{widetext}
where \(H^{\mu\nu}_{NP_{11}}\) is the DVCS virtual Compton tensor for the transition, defined as
\begin{align}
    \nonumber H_{NP_{11}}^{\mu\nu} = &-i\int\dd^4y\, e^{-iqy} \\
    &\times \Braket{R(p_R,s_R)| T[J^\mu_{\text{em}}(0) J^\nu_{\text{em}}] |N(p,s_N)} \, ,
\end{align}
and the pion emission is governed by a pseudovector vertex, consistent with chiral symmetry.

\begin{figure}
    \centering
    \includegraphics[width=0.9\linewidth]{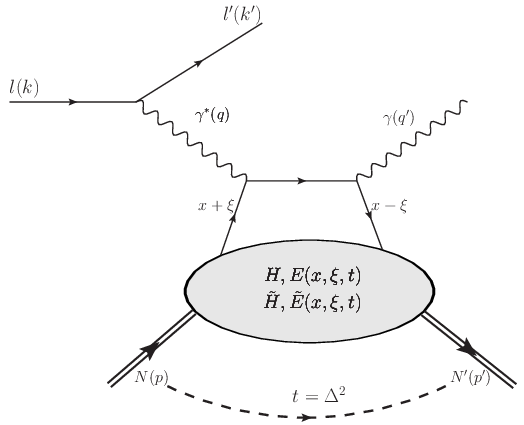}
    \caption{Handbag diagram of the DVCS process in the DGLAP region \(x\in[\xi,1]\), corresponding to the scattering on a quark~\cite{fig:handbag}.}
    \label{fig:handbag}
\end{figure}

In the Bj{\"o}rken limit (\(Q^2\to\infty, x_B\) constant) to leading twist (twist-2) the DVCS amplitude is then calculated from the diagram in Figure~\ref{fig:handbag}, which we parametrise in terms of 4 transition GPDs.

For the evaluation of these diagrams it is useful to introduce the light-like vectors \(\tilde{p}\), \(n\):
\begin{align}
    \bar{P}^\mu = \dfrac{p^\mu + p_R^\mu}{2} &= \tilde{p}^\mu + \dfrac{p^\mu + p_R^\mu}{4}n^\mu \, , \\
    q^\mu &= (-2\xi')\tilde{p}^\mu + \dfrac{Q^2}{4\xi'}n^\mu \, ,
\end{align}
with
\begin{gather}
    \xi' = \frac{q\cdot\bar{P}}{2\bar{P}^2} \left[-1 + \sqrt{1 + \frac{Q^2\bar{P}^2}{(q\cdot\bar{P})^2}}\right].
\end{gather}

In the Bj{\"o}rken limit
\begin{gather}
    \xi' \approx \xi \to \frac{x_B/2}{1-x_B/2}
\end{gather}
and the leading twist tensor can be expressed as
\begin{widetext}
\begin{eqnarray}
    \nonumber H^{\mu\nu}_{NP_{11}}=\frac{1}{2}(-g^{\mu\nu}_\perp)\int_{-1}^1\dd x\left[\frac{1}{x-\xi+i\epsilon}+\frac{1}{x+\xi-i\epsilon}\right]\int\frac{\dd\lambda}{2\pi}e^{i\lambda x} \sum_q e_q^2 \braket{P_{11}(p_R,s_R)|\bar{q}\left(-\frac{\lambda n}{2}\right)\gamma\cdot nq \left(\frac{\lambda n}{2}\right)|N(p,s_N)} \\
    \nonumber + \frac{i}{2}(\varepsilon^{\mu\nu}_\perp) \int_{-1}^1\dd x\left[\frac{1}{x-\xi+i\epsilon} - \frac{1}{x+\xi-i\epsilon}\right]\int\frac{\dd\lambda}{2\pi}e^{i\lambda x} \sum_q e_q^2 \braket{P_{11}(p_R,s_R)|\bar{q}\left(-\frac{\lambda n}{2}\right)\gamma\cdot n\gamma^5 q \left(\frac{\lambda n}{2}\right)|N(p,s_N)} \, , \\
\end{eqnarray}
\end{widetext}
with transverse projectors
\begin{align}
    \nonumber -g_\perp^{\mu\nu} &= -g^{\mu\nu} + \tilde{p}^\mu n^\nu + \tilde{p}^\nu n^\mu, \\
    \varepsilon_\perp^{\mu\nu} &= \varepsilon^{\mu\nu\alpha\beta} n_\alpha  \tilde{p}_\beta \, .
\end{align}

This is the convolution integral between a perturbative kernel over the average quark momentum \(x\) representing quark propagators between the photon legs of the handbag diagram, and a non-perturbative quark bilinear operator along the light-cone direction \(n^\mu\), which we will parametrize in terms of transition GPDs.
\begin{widetext}
\begin{align}
    \nonumber \int\frac{\dd\lambda}{2\pi}e^{i\lambda x} \sum_q e_q^2 & \braket{P_{11}(p_R,s_R)|\bar{q}\left(-\frac{\lambda n}{2}\right)\gamma\cdot nq \left(\frac{\lambda n}{2}\right)|N(p,s_N)} \\
    &= \bar{R}(p_R,s_R) \left\{ H_1^{NP_{11}}(x,\xi,\Delta^2) \left(n^\nu - \frac{n\cdot\Delta}{\Delta^2}\Delta^\nu\right)\gamma_\nu + H_2^{NP_{11}}(x,\xi,\Delta^2) \frac{i\sigma_{\nu\kappa}n^\nu\Delta^\kappa}{M_R + M_N} \right\} N(p,s_N) \, , \\
    \nonumber \int\frac{\dd\lambda}{2\pi}e^{i\lambda x} \sum_q e_q^2 & \braket{P_{11}(p_R,s_R)|\bar{q}\left(-\frac{\lambda n}{2}\right)\gamma\cdot nq \left(\frac{\lambda n}{2}\right)|N(p,s_N)} \\
    &= \bar{R}(p_R,s_R) \left\{ \widetilde{H}_1^{NP_{11}}(x,\xi,\Delta^2)\gamma\cdot n\gamma_5 + \widetilde{H}_2^{NP_{11}}(x,\xi,\Delta^2) \frac{\Delta\cdot n}{M_R + M_N}\gamma^5 \right\} N(p,s_N) \, ,
\end{align}
\end{widetext}
for the vector and axial-vector bilinears, respectively.

The transition GPDs are constructed from a sum of transition quark GPDs, weighted by the quadratic quark charges
\begin{align}
    H_i^{pP_{11}} &= \frac{4}{9}H_i^{u,pP_{11}} + \frac{1}{9}H_i^{d,pP_{11}} \, , \\
    H_i^{nP_{11}} &= \frac{1}{9}H_i^{u,nP_{11}} + \frac{4}{9}H_i^{d,nP_{11}} \, ,
\end{align}

We denote the vector GPDs by \(H_1\) and \(H_2\), and the axial-vector GPDs by \(\widetilde{H}_1\) and \(\widetilde{H}_2\). In each sector, the GPD multiplying the \(\gamma^\mu\) (\(\gamma^\mu\gamma^5\)) structure is sometimes referred to as an \(H\)-type GPD, while the one multiplying the Pauli-type structure is referred to as an \(E\)-type GPD.

\subsection{Diagonal Process} 
\label{sec:diagonalAmplitude}
The scattering amplitude for the process where a target nucleon emits a pion, is then struck by a virtual photon, emits a real photon, and remains a nucleon is given by
\begin{widetext}
\begin{equation}
    i\mathcal{M} = \frac{e\,g_f\,g_A\,\tau^a}{2f_\pi} \, \bar{u}_e(k_e',s') \, \gamma_\mu \, u_e(k_e,s) \, \varepsilon^*_\nu(q',\lambda') \, \bar{u}_N(p_R-k_\pi,s_N') \, H^{\mu\nu}_{NN} \, S_f(p - k_\pi,M_N') \, \slashed{k}_\pi\gamma^5 \, F_{N\!N\pi}(|\vv{k_\pi}|;\Lambda) \, u_N(p,s_N) \, ,
\end{equation}
\end{widetext}
where \(H^{\mu\nu}_{NN}\) is the diagonal DVCS Compton tensor, defined as
\begin{align}
    \nonumber & H_{NN}^{\mu\nu} = -i\int\dd^4y\, e^{-iqy} \\
    &\times \Braket{N(p_R-k_\pi,s_N')| T[J^\mu_{\text{em}}(0) J^\nu_{\text{em}}] |N(p-k_\pi,s_N)} \, ,
\end{align}
and the pion emission is governed by the same pseudovector vertex factor, dressed by a dipole form factor,
\begin{gather}
    F_{N\!N\pi}(|\vv{k_\pi}|;\Lambda) = \left( \frac{\Lambda^2}{|\vv{k_\pi}|^2 + \Lambda^2}\right)^2 \, .
\end{gather}

Because the strong pion emission process occurs before the hard scattering, the pion three-momentum is not kinematically tied to the Roper two-body decay value and can run large for the boosted \(N\pi\) system, especially for large \(Q^2\) and \(-t\). This enhances the off-shellness of the intermediate nucleon and enters the amplitude directly through the pseudovector coupling. In the absence of a damping factor, the resulting contribution is therefore sensitive to off-shell extrapolation of the effective \(N\pi\) vertex. We adopt \(\Lambda=0.7~\mathrm{GeV}\), consistent with typical cutoff scales used in meson-exchange descriptions of \(N\pi\) dynamics, such as the Sato–Lee model~\cite{SatoLee1996}.

Since the Roper has the same quantum numbers as a nucleon, the parametrisation in terms of GPDs closely mirrors that of the transition.

\begin{align}
    \bar{P}^\mu = \dfrac{p^\mu + p_R^\mu - 2k_\pi^\mu}{2} &= \tilde{p}^\mu + \dfrac{p^\mu + p_R^\mu - 2k_\pi^\mu}{4}n^\mu \, , \\
    q^\mu &= (-2\xi')\tilde{p}^\mu + \dfrac{Q^2}{4\xi'}n^\mu \, ,
\end{align}

\begin{widetext}
\begin{align}
    \nonumber H^{\mu\nu}_{NN} &=\frac{1}{2}(-g^{\mu\nu}_\perp)\int_{-1}^1\dd x\left[\frac{1}{x-\xi+i\epsilon}+\frac{1}{x+\xi-i\epsilon}\right] \\
        & \qquad\qquad \times\int\frac{\dd\lambda}{2\pi}e^{i\lambda x} \sum_q e_q^2 \braket{N(p_R-k_\pi,s_N')|\bar{q}\left(-\frac{\lambda n}{2}\right)\gamma\cdot nq \left(\frac{\lambda n}{2}\right)|N(p-k_\pi,s_N)} \\
        \nonumber &+ \frac{i}{2}(\varepsilon^{\mu\nu}_\perp) \int_{-1}^1\dd x\left[\frac{1}{x-\xi+i\epsilon} - \frac{1}{x+\xi-i\epsilon}\right] \\
        & \qquad\qquad \times \int\frac{\dd\lambda}{2\pi}e^{i\lambda x} \sum_q e_q^2 \braket{N(p_R-k_\pi,s_N')|\bar{q}\left(-\frac{\lambda n}{2}\right)\gamma\cdot n\gamma^5 q \left(\frac{\lambda n}{2}\right)|N(p-k_\pi,s_N)} \, . \\
    \nonumber \int\frac{\dd\lambda}{2\pi}e^{i\lambda x} \sum_q e_q^2 & \braket{N(p_R,s_R)|\bar{q}\left(-\frac{\lambda n}{2}\right)\gamma\cdot nq \left(\frac{\lambda n}{2}\right)|N(p,s_N)} \\
        &= \bar{N}(p_R-k_\pi,s_N') \left\{ H_1^{NN}(x,\xi,\Delta^2) \left(n^\nu - \frac{n\cdot\Delta}{\Delta^2}\Delta^\nu\right)\gamma_\nu + H_2^{NN}(x,\xi,\Delta^2) \frac{i\sigma_{\nu\kappa}n^\nu\Delta^\kappa}{M_R + M_N} \right\} N(p-k_\pi,s_N), \\
    \nonumber \int\frac{\dd\lambda}{2\pi}e^{i\lambda x} \sum_q e_q^2 & \braket{N(p_R,s_R)|\bar{q}\left(-\frac{\lambda n}{2}\right)\gamma\cdot nq \left(\frac{\lambda n}{2}\right)|N(p,s_N)} \\
        &= \bar{N}(p_R-k_\pi,s_N') \left\{ \tilde{H}_1^{NN}(x,\xi,\Delta^2)\gamma\cdot n\gamma_5 + \tilde{H}_2^{NN}(x,\xi,\Delta^2) \frac{\Delta\cdot n}{M_R + M_N} \right\} N(p-k_\pi,s_N) \, ,
\end{align}
\end{widetext}
where
\begin{align}
    H_i^{pp} &= \frac{4}{9}H_i^{u,pp} + \frac{1}{9}H_i^{d,pp} \\
    H_i^{nn} &= \frac{1}{9}H_i^{u,nn} + \frac{4}{9}H_i^{d,nn} \, .
\end{align}

\subsection{Final-State Interactions} 
\label{sec:FSI}
For the diagonal case, where \(N\pi\) rescattering between the final-state nucleon and pion is not accounted for by an explicit Roper propagator, we enforce Watson's 
Theorem~\cite{Watson:1952ji} via a phase factor
\begin{gather}
    \mathcal{M}_{\mathrm{FSI}}=\mathcal{M}_{\mathrm{bare}} \times e^{i\delta_{N\pi}(W)} \, ,
\end{gather}
where \(\delta_{N\pi}(W) = \arg\!\left(W^2 - M_R^2 + i M_R \Gamma_R\right)\) is the \(P_{11} N\pi\) phase shift in a single-resonance approximation and we neglect the energy-dependence of the width, instead using the PDG value, \(\Gamma = 0.350\;\mathrm{GeV}\).

\section{Model for the GPDs} 
\label{sec:GPDModels}
We expand upon the minimal model for the Nucleon-to-Roper transition GPDs presented in Ref.~\cite{Semenov-Tian-Shansky:2023bsy} by using a light-cone double-distribution framework, with a Regge-inspired \(t\)-dependence. Whilst the minimal model ensures the fixing of the first Mellin moments to the transition form factors, support for \(x\in[0,1]\), and a reasonable valence-like shape, it introduces no skewness-dependence in the GPD, assumes a completely factorised \(t\)-dependence, and does not enforce polynomiality of higher Mellin moments.

The use of a double distribution representation ensures the Mellin moments in \(x\) are exact polynomials in \(\xi\) of the correct order (up to a here-neglected \(D\)-term) \cite{Guidal:2004nd}. Additionally, a non-trivial skewness-dependence enters the GPD via the profile function and not just from the handbag kernel. This is vitally important for understanding interference patterns with the Bethe--Heitler amplitude, controlling the relative contributions of the DGLAP and ERBL regions, and performing scans over broader \(x\) ranges than those accessible at CLAS12.

In Ref.~\cite{Semenov-Tian-Shansky:2023bsy} the transverse profile of the transition GPDs is \(x\)-independent; that is, the same \(t\)-slope applies for all \(x\). Using a Regge-style \(t\)-dependence inside the double distribution encodes a steeper roll-off for small values of \(\beta\) (predominantly small \(x\) in the DGLAP region). This reflects the empirically-supported idea that small-\(x\) physics is more peripheral than large-\(x\) and, by correlating with \(t\), gives a more realistic impact-parameter picture of the transition.

While there are no direct valence quark PDFs for the transition, we can tie the small-\(x\) behaviour at \(t=0\) to Regge trajectories extracted from Roper electroproduction and enforce quark counting laws for \(x\to1\). This model is then internally consistent with both hadronic Regge phenomenology and hard-scattering counting rules.

While this model incorporates some phenomenological improvements, for the kinematics investigated here it provides little deviation from the model used in Ref.~\cite{Semenov-Tian-Shansky:2023bsy}. The main benefit is the easy application of the same principles to the Nucleon GPD model.

Let us discuss in detail the formulation of this model. Our GPDs obey the basic form
\begin{eqnarray}
    \nonumber \label{eq:genGPD}
    H\!\left(x,\xi,t\right)=\int_{-1}^1\dd\beta \int_{-1+\left|\beta\right|}^{1-\left|\beta\right|}\dd\alpha \, \delta\!\left(x-\beta-\xi\alpha\right) \\
    \times \pi\!\left(\beta,\alpha\right) \, q\!\left(\beta\right) \, e^{b\left(1-\beta\right)t} \, ,
\end{eqnarray}
where \(q(\beta)\) is a valence parton distribution function, \(\pi\!\left(\beta,\alpha\right)\) is a profile function and \(e^{b\left(1-\beta\right)t}\) is the Regge-style \(t\)-dependence, with profile parameter \(b\).

We restrict ourselves to valence contributions and take \(q(\beta<0)=0\), so anti-quark contributions are neglected. We also omit an explicit \(D\)-term; its effect on the observables considered here is expected to be small at our kinematics.

In this work we choose \(b^u_{H_1} = 1.0\), \(b^d_{H_1} = 1.2\), \(b^{u(d)}_{\widetilde H_1}=1.1\), with broader slopes \(b^u_{H_2} = 1.8\), and \(b^d_{H_2} = 2.3\) for the impact-parameter-style slopes, for both the transition and diagonal GPDs. For the transition GPD, \(\widetilde H_2^{u(d)}\), we use \(b^{u(d)}_{\widetilde H_2}= 1.8(2.2)\), respectively and directly set \(\widetilde H_2 = 0\) for the diagonal case, since it is negligible. This is because it contributes to the amplitude through helicity-flip structures which, in unpolarised cross sections and many beam-spin observables, typically occur with kinematic powers of \(\xi\) and \(-t/4M_N^2\). Therefore at moderate skewness and momentum transfer, \(\widetilde{H}_2\) is subleading compared to \(\widetilde{H}_1\) and the unpolarised GPDs. \(\widetilde{H}_2\) does have a well-known pion-pole contribution \(\propto 1/(m_\pi^2-t)\) \cite{DIEHL200341}, but for physical DVCS with \(t<0\) this pole enhancement rapidly dies for \(-t\gg m_\pi^2\).
To confirm this we extend the model to \(\widetilde{H}_2\),  finding that the diagonal and combined cross sections are suppressed by \(<2\%\) at \(-t=0.35~\mathrm{GeV}\), with no visible effect by \(-t=1.5~\mathrm{GeV}\). We consider this negligible and therefore set \(\widetilde{H}_2\) to zero hereinafter.

We use the Radyushkin profile function \cite{RADYUSHKIN1996417, PhysRevD.59.014030, RADYUSHKIN1996333}
\begin{equation}
  \pi\!\left(\beta,\alpha\right) = \frac{3}{4}\frac{\left(1-\beta\right)^2-\alpha^2}{\left(1-\beta\right)^3}
\end{equation}
and quark PDF parametrizations
\begin{align}
    \nonumber q_u(\beta) &= N_u\beta^{-\alpha_0} (1 - \beta)^{n_u} \, , \\
    \nonumber q_d(\beta) &= N_d\beta^{-\alpha_0} (1 - \beta)^{n_d} \, , \\
    \nonumber q_{u,A}(\beta) &= N_{u,A}\beta^{-\alpha_0} (1 - \beta)^{n_{u,A}} \, , \\
    q_{d,A}(\beta) &= N_{d,A}\beta^{-\alpha_0} (1 - \beta)^{n_{d,A}} \, ,
\end{align}
with Regge intercept \(\alpha_0\) and normalization constants \(N\).

We choose \(\alpha_0=0.5\) for the Regge intercept, \(n_u=3\), \(n_d=4\), \(n_{u,A}=3\) and \(n_{d,A}=5\) for the large-\(x\) fall-off in the diagonal case. For the transition case we use the same values of \(\alpha_0,\, n_u,\, n_d\), but do not use a separate shape for the PDFs that enter the axial GPDs.

For the diagonal GPDs, the normalizations are chosen to match the quark charges and anomalous moments at \(t=0\), i.e. \(\int_0^1\dd\beta\,q_u=2\), \(\int_0^1\dd\beta\,q_d=1\), and \(\int_0^1\dd\beta\,H_2^{u(d)}(\beta,0,0)=\kappa_{u(d)}\).  The axial inputs are normalised to \(\Delta u\) and \(\Delta d\). For the transition GPDs, the normalisations are chosen such that the first Mellin moments of the GPDs correctly reproduce the form factors \(F_{1,2}(t)\) for the proton and neutron from MAID2008 helicity amplitude data for all \(t\) in our kinematical region. The transition axial GPDs are normalised to fix the first Mellin moments to \(G_A(t)\) and \(G_P(t)\), respectively, with a pion-pole form for \(G_P\).

Finally, we obtain the charge-squared-weighted flavor GPDs for the proton(neutron) by combining the up- and down-quark GPDs in the correct combinations, assuming isospin symmetry
\begin{align} \label{eq:nucleonGPD}
    \nonumber H_{1(2)}^p\left(x,\xi,t\right) &= \frac{4}{9}H_{1(2)}^u\!\left(x,\xi,t\right) + \frac{1}{9}H_{1(2)}^d\!\left(x,\xi,t\right) \, , \\
    H_{1(2)}^n\left(x,\xi,t\right) &= \frac{1}{9}H_{1(2)}^u\!\left(x,\xi,t\right) + \frac{4}{9}H_{1(2)}^d\!\left(x,\xi,t\right) \, ,
\end{align}
and similarly for the polarised GPDs.

This parametrisation allows the direct evaluation of Compton form factors through convolution with the hard scattering kernel over \(x\). Compared to the minimal model of Semenov {\em et al.}~\cite{Semenov-Tian-Shansky:2023bsy}, which uses a purely valence-like \(x\)-dependence with factorised \(t\) and trivial skewness structure, the present DD-based parametrisation enforces the correct polynomiality pattern of Mellin moments (up to neglected \(D\)-term contributions \cite{PhysRevD.87.096017}), is dominated by valence PDFs with simple functional forms, generates non-trivial skewness dependence via the profile function, and correlates the transverse profile with \(x\) inspired by Regge phenomenology. The normalisations are fixed to reproduce diagonal charges and anomalous moments at \(t=0\), and MAID2008(MAID2007) transition form 
factors~\cite{MAID} as functions of \(t\), ensuring consistency with existing electroproduction data on the proton(neutron), while remaining computationally tractable.

\section{Results} \label{sec:analysis}
\begin{figure*}
\begin{subfigure}{0.49\textwidth}
    \centering
    \includegraphics[width=\linewidth]{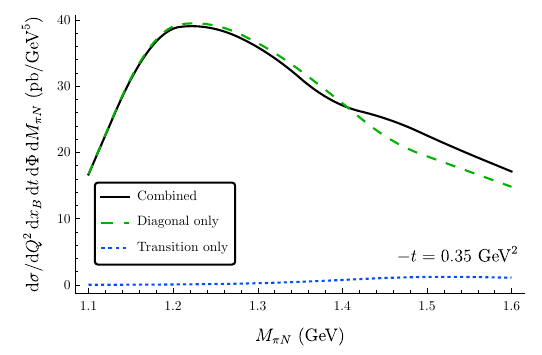}
\end{subfigure}
\begin{subfigure}{0.49\textwidth}
    \centering
    \includegraphics[width=\linewidth]{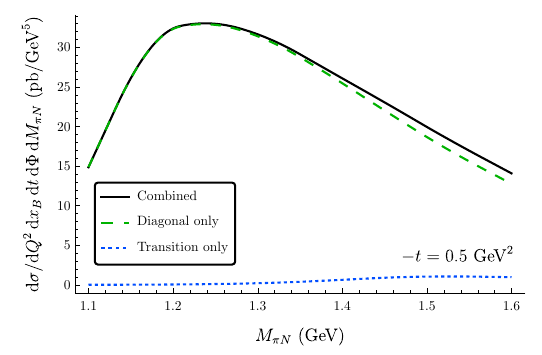}
\end{subfigure}
\begin{subfigure}{0.49\textwidth}
    \centering
    \includegraphics[width=\linewidth]{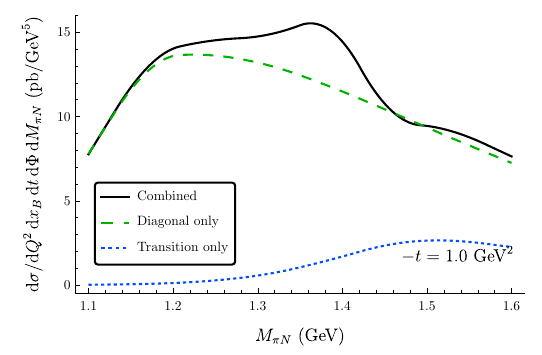}
\end{subfigure}
\begin{subfigure}{0.49\textwidth}
    \centering
    \includegraphics[width=\linewidth]{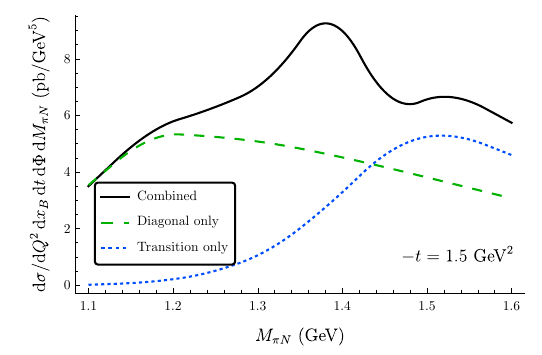}
\end{subfigure}
    \caption{Differential cross section for the DVCS process \(p^+ e^- \to p^+ \pi^0 e^-\gamma\) evaluated at beam energy \(E_e=10.6~\mathrm{GeV}\), photon virtuality \(Q^2=2.3~\mathrm{GeV}^2\), Bj{\"o}rken-\(x\) \(x_B=0.2\), scattering plane angle \(\Phi=\pi/2\), and various momentum transfers \(-t\). Integrated over the pion decay solid angle \(\Omega^*_\pi\) with the cut \(M_{\pi\gamma}>1~\mathrm{GeV}\) applied.}
    \label{fig:crossSectionNeutral}
\end{figure*}

To facilitate comparison with CLAS12 analyses from JLab, we choose kinematics close to those expected in forthcoming data. We begin by considering the case where the hadronic final state consists of a proton and a neutral pion. In Fig.~\ref{fig:crossSectionNeutral} we show the dependence of the differential cross section on the invariant mass of the \(N\pi\) system for the reaction \(p^+ e^- \to p^+ \pi^0 e^-\gamma\), subject to the cut \(M_{\pi\gamma}>1~\mathrm{GeV}\), which is chosen in order to minimise the impact of the \(p^+e^-\to\rho^+e^-\to\gamma\pi e^-\) background channel. To be clear on the terminology used, we refer to the subprocess where the hard scattering yields a nucleon as \textbf{``diagonal''}, while that in which the hard process produces an excited baryon is labelled \textbf{``transition''}. We plot the individual contributions for both the diagonal and transition subprocesses, as well as the combined cross section for the total DVCS process. 

We first note that the contribution from the transition to the excited baryon state is quite small relative to the diagonal process at low \(-t\) but increases with larger \(-t\). This effect is primarily driven by a suppression in the magnitude of the diagonal process along with a small enhancement in the transition term. The interference created in the total process (\(\mathcal{M}_\mathrm{diag}\mathcal{M}_\mathrm{trans}^* + \mathcal{M}_\mathrm{diag}^*\mathcal{M}_\mathrm{trans}\)) follows the same pattern, becoming sizeable for \(-t>1~\mathrm{GeV}\). This is unsurprising, given that the dipole form factor, dressing the pion emission for the diagonal case, suppresses pions with large three-momentum, which in our kinematics is correlated with larger \(-t\) in the diagonal mechanism. Additionally, the diagonal GPDs consistently decay for large \(-t\), whereas the transition GPDs rise to a peak before decaying (at around \(-t=1.74~\mathrm{GeV}\) for the proton and \(-t=2.26~\mathrm{GeV}\) for the neutron). This suggests measurable sensitivity to the transition GPDs, which would therefore be best observed at momentum transfers near these peaks in experimental DVCS observations.

\begin{figure*}
\begin{subfigure}{0.49\textwidth}
    \centering
    \includegraphics[width=\linewidth]{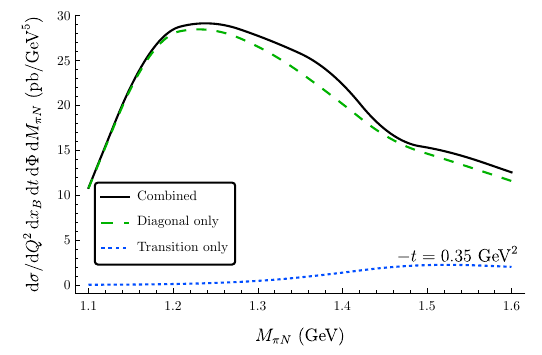}
\end{subfigure}
\begin{subfigure}{0.49\textwidth}
    \centering
    \includegraphics[width=\linewidth]{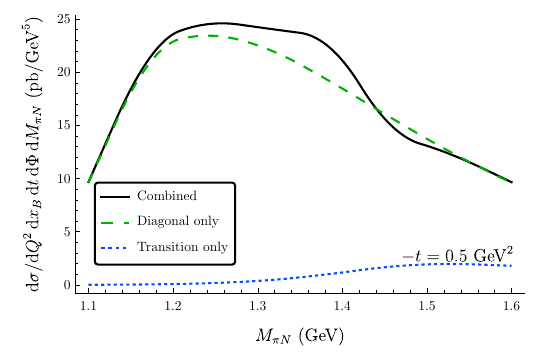}
\end{subfigure}
\begin{subfigure}{0.49\textwidth}
    \centering
    \includegraphics[width=\linewidth]{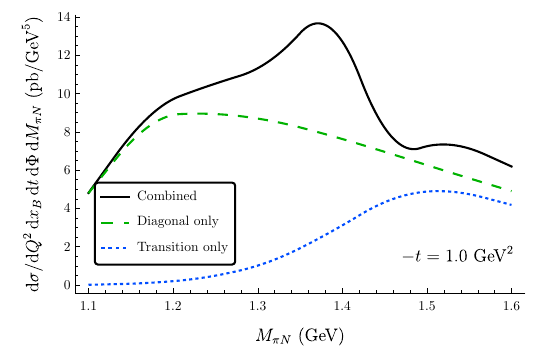}
\end{subfigure}
\begin{subfigure}{0.49\textwidth}
    \centering
    \includegraphics[width=\linewidth]{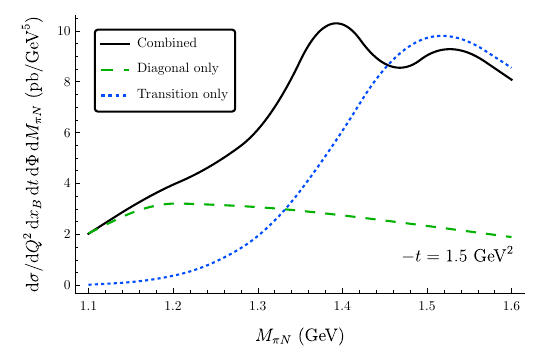}
\end{subfigure}
    \caption{Differential cross section for the DVCS process \(p^+ e^- \to n \pi^+ e^- \gamma\) evaluated at beam energy \(E_e=10.6~\mathrm{GeV}\), photon virtuality \(Q^2=2.3~\mathrm{GeV}^2\), Bj{\"o}rken-\(x\) \(x_B=0.2\), scattering plane angle \(\Phi=\pi/2\), and various momentum transfers \(-t\). Integrated over the pion decay solid angle \(\Omega^*_\pi\) with the cut \(M_{\pi\gamma}>1~\mathrm{GeV}\) applied.}
    \label{fig:crossSectionCharged}
\end{figure*}

In Fig.~\ref{fig:crossSectionCharged} we show the results under the same kinematic conditions, but for the creation of a charged pion via \(p^+ e^- \to n \pi^+ e^-\gamma\). We find a more pronounced impact on the total cross section than for the neutral case with the transition term making a significant contribution to the total cross section from lower \(-t\). This is a consequence of the smaller diagonal GPDs for the neutron than the proton, consistent with the smaller charge-weighted Compton amplitude on the neutron. This suppression arises from the isospin structure of the neutron GPDs, whereby the dominant \(u\)-quark contribution enters with a reduced charge weight in the neutron compared to the proton as per Eq. \ref{eq:nucleonGPD}.

Our treatment of the transition subprocess closely follows the formalism developed in Ref.~\cite{Semenov-Tian-Shansky:2023bsy}, and our results are consistent with its main qualitative conclusions. In particular, we find the same overall hierarchy and kinematic trends for the transition contribution: the transition signal is strongly channel- and $t$-dependent, becoming relatively more important at larger $-t$ and in channels where the diagonal amplitude is reduced by charge/flavor weighting.

\section{Conclusions and Outlook} 
\label{sec:conclusion}
We have studied the exclusive process $eN\to e\gamma N\pi$ under kinematic conditions relevant to forthcoming CLAS12 measurements, with the specific aim of assessing how intermediate-state pion emission can modify the observed DVCS-like signal and how sensitive the reaction may be to $N\to N^*$ transition GPDs. We constructed the DVCS amplitude for both a diagonal-like subprocess and a transition subprocess proceeding through the Roper region, and combined these with the pion-emission mechanism that accompanies the hard scattering. Within the model used here (leading-twist handbag factorization for the Compton subprocess, a phenomenological transition-GPD model, and a regulated $NN\pi$ vertex), we obtained quantitative estimates of the size and kinematic dependence of these effects.

Our numerical results indicate four main qualitative outcomes.
\begin{itemize} 
	\item First, pion-emission contributions can produce non-negligible modifications to the $eN\to e\gamma N\pi$ differential cross section in realistic kinematics. Consequently, they must be included in precision studies. 
	\item Second, the relative importance of diagonal and transition mechanisms depends strongly on the channel and on $-t$: for neutral-pion production on the proton the diagonal term tends to dominate at smaller $-t$, while the transition contribution grows in relative importance as $-t$ increases; for charged-pion production the transition contribution can become competitive already at more moderate $-t$, consistent with the different charge weighting in the neutron diagonal Compton amplitude under isospin symmetry.
	\item Third, because the observable rate receives interference between diagonal and transition amplitudes, there exist kinematic regions where sensitivity to the transition GPDs is enhanced beyond what one would infer from the transition contribution alone. This supports the broader motivation that exclusive DVCS-like production of resonance final states can provide experimentally useful access to transition GPD information.
	\item Finally, in terms of distinguishing the QCD origin of the Roper resonance, the relative contributions of diagonal and transition amplitudes do appear to offer considerable promise, as the latter subprocess would be parametrically-reduced or have a distinct kinematic dependence if the Roper were dynamically generated.
\end{itemize}

There are several clear limitations of the present study that also define the most important next steps. Most notably, we have not included the Bethe--Heitler contribution (and its interference with DVCS), which is essential for direct comparison to experimentally measured beam-spin and charge asymmetries. Extending the present framework to include the full Bethe--Heitler amplitude for the $2\to4$ process, including the relevant experimental acceptance cuts and radiative backgrounds, is therefore a priority. In addition, it will be important to broaden the resonance content beyond a single effective Roper contribution (e.g.\ inclusion of nearby $N^*$ and $\Delta^*$ states in the same analysis) and to test the robustness of the predicted transition sensitivity against reasonable variations of the transition-GPD model (profile functions, Regge slopes, and normalization constraints from transition form factors).

Looking forward, the most direct application of this work is to provide a controlled signal model for CLAS12 analyses of $eN\to e\gamma N\pi$ in the resonance region. With the Bethe--Heitler process included, one can identify optimized observables that suppress diagonal contributions and enhance the interference sensitivity to transition GPDs, making resonance-tagged DVCS a practical tool for mapping transition GPDs and for testing hypotheses about the dynamical origins of the Roper resonance.

Although the present study targets CLAS12-like kinematics (valence-dominated $x_B$), this GPD model remains instructive when extrapolated to the smaller-$x$ regime accessible at an EIC. At a structural level, the double-distribution representation provides a controlled implementation of polynomiality (up to a neglected $D$-term) and introduces genuine skewness dependence through the profile function, features that remain required at small $x$. Moreover, the Regge-motivated small-$x$ behavior embedded in the input distributions suggests that the present ansatz can serve as a practical interpolation framework between fixed-target and collider kinematics.

Quantitatively, however, the predictive power of a valence-constrained model decreases as $x$ becomes small. In particular, sea-quark and gluon GPDs become increasingly important, and the larger $Q^2$ lever arm at an EIC implies substantially longer perturbative evolution, so gluon mixing and higher-order evolution effects can materially alter both normalization and skewness patterns. At sufficiently small $x$, additional dynamics beyond a simple Regge extrapolation (e.g. saturation effects) may also become relevant. For these reasons, we view the present modelling as qualitatively informative down to moderately small $x$ where collinear factorization and leading-twist evolution remain under control, while robust EIC predictions will require an extended framework with explicit sea/gluon components, consistent (N)LO evolution, and confrontation with small-$x$ constraints.

\section*{Acknowledgements} This work was supported by Adelaide University through the Centre for the Subatomic Structure of Matter and by the Australian Research Council through Discovery Project DP230101791 (AWT).

\bibliography{refs}

\end{document}